\documentclass[letterpaper]{article}
\usepackage{aaai22}  
\usepackage{times}  
\usepackage{helvet} 
\usepackage{courier}  
\usepackage[hyphens]{url}  
\usepackage{graphicx} 
\usepackage{booktabs}
\usepackage{natbib}  
\usepackage{caption} 
\usepackage{multicol}
\usepackage{subcaption}

\usepackage{cleveref}
\usepackage{multicol}
\usepackage{array,multirow}
\usepackage{subcaption}
\usepackage{capt-of}
\usepackage{tablefootnote}

\usepackage[export]{adjustbox}

\begin{document}

\title{Examining the Production of Co-active Channels on YouTube and BitChute} 

\author{Matthew C. Childs and Benjamin D. Horne}
\affiliations{School of Information Sciences, The University of Tennessee, Knoxville, Knoxville, TN, USA\\mchilds3@vols.utk.edu, bhorne6@utk.edu}

\maketitle

\begin{abstract}
A concern among content moderation researchers is that hard moderation measures, such as banning content producers, will push users to more extreme information environments. Research in this area is still new, but predominately focuses on one-way migration (from mainstream to alt-tech) due to this concern. However, content producers on alt-tech social media platforms are not always banned users from mainstream platforms, instead they may be co-active across platforms. We explore co-activity on two such platforms: YouTube and BitChute. Specifically, we describe differences in video production across 27 co-active channels. We find that the majority of channels use significantly more moral and political words in their video titles on BitChute than in their video titles on YouTube. However, the reasoning for this shift seems to be different across channels. In some cases, we find that channels produce videos on different sets of topics across the platforms, often producing content on BitChute that would likely be moderated on YouTube. In rare cases, we find video titles of the same video change across the platforms. Overall, there is not a consistent trend across co-active channels in our sample, suggesting that the production on alt-tech social media platforms does not fit a single narrative.
\end{abstract}

\section{Introduction} 
As disinformation, misinformation, and hate speech continue to infect social media platforms, researchers continue to focus on methods to limit the visibility and consumption of bad content. These efforts have lead to a wide range of potential solutions, from banning repeat offenders to quarantining communities to placing warning labels on content. Some of these solutions are automated by machine learning while others are completely human-powered solutions. 

While progress on content moderation has been made, a primary concern among researchers is the impact of hard moderation measures, such as banning users from a social media platform. Specifically, researchers are concerned that deplatforming efforts might end up pushing users to more toxic environments, perhaps mitigating the pro-social benefit of banning users in the first place \cite{russo2022spillover}. Due to this concern, recent research has focused on movement from one platform to another, in which a content producer is banned from or intervened with on one platform, leaves, and starts producing content on another. These studies have provided evidence that some users do indeed migrate from one platform to another after a moderation event, and subsequently can increase their toxic behaviors \cite{ali2021understanding, horta2021platform}. Although, importantly, the audiences of the toxic content producers decreases after the movement \cite{ali2021understanding}, and in some cases the activity of migrants themselves decreases after the movement \cite{horta2021platform}. There are some additional complexities to this evidence. For example, in the case of toxic Reddit communities, subreddits where the banned community migrants went to did not inherit the toxicity of the banned communities \cite{chandrasekharan2017you}. Hence, a generalized understanding of user migrations and their impacts remains unclear. 

Under this umbrella of work, the primary focus has been one-way migration (migration from one platform to another) usually as a result of being deplatformed. However, one-way migration is only one of several potential side-effects of content moderation. For many content producers, it is not necessarily a choice of being on one platform or the other, but instead content producers may operate accounts across the both mainstream and alternative platforms. For example, content producers who have not been banned but still perceive a slight from content moderation, from actions such as being demonetized, having warning labels placed on their content, or the banning of another influencer within a specific toxic community, may begin producing content in an alternative space without leaving the mainstream platform. The primary concern here is not just that users may end up in more toxic environments, but that they will end up in more toxic environments while still having the affordances of mainstream platforms, possibly creating bridges to bring consumers with them to alternative spaces. Recently, work by \citet{russo2022spillover} provided evidence that ``\textit{spillovers}'' of antisocial behavior can occur between alternative and mainstream platforms through co-active users. Specifically, they showed that co-active users were more antisocial on the mainstream platform than other users of the mainstream platform, and that this effect escalated over time with exposure to the fringe platform. Outside of this work, which focused on toxic Reddit communities co-activity with fringe platforms, co-active behavior has been relatively unexplored.

To this end, we describe co-active behavior across two different mainstream and alternative platforms: YouTube and BitChute (a fringe alternative to YouTube \cite{trujillo2020bitchute}). Specifically, in this short paper we ask two basic questions:
\begin{itemize}
    \item \textbf{RQ1:} How do co-active content producers differ in their production across YouTube and BitChute?
    \item \textbf{RQ2:} How do co-active content producers change content toxicity and language across YouTube and BitChute?
\end{itemize}

To explore these questions, we collect and analyze a unique dataset of videos and comment metadata from 27 matched channels on YouTube and BitChute between October 2020 and October 2021. Using this dataset, we found that many content producers do not use significantly different language in their video titles across YouTube and BitChute, despite many videos not being exact copies of each other. However, there are some notable exceptions to this generalization. 

In particular, we found that the majority of channels use more moral (e.g. wrong, honor, deserve, judge) and political (e.g. united states, govern, congress, senate) words in their video titles on BitChute than in their video titles on YouTube. The reasoning for this shift seems to be different across channels. In some cases, we find that channels produce videos on different sets of topics across the platforms, often producing content on BitChute that would likely be moderated on YouTube. In other cases, this shift was due to the timing of videos produced. For example, a channel may produce more videos on YouTube early in the time frame, discussing a specific event, and then produce more videos on BitChute later in the time frame, discussing a new set of events. Broadly, these inconsistencies suggest that the production on and movement to alt-tech social media platforms does not fit a single narrative.

\section{Data and Methods}
Our methods are loosely inspired by \citet{horta2021platform}, in which the authors examine toxic communities after those communities move to new platforms after moderation measures from their original platform. In our case, we are examining concurrent behavior, rather than before and after an intervention. Further, our study is focused on video content creators rather than members of discussion communities. These differences change some of our methods of analysis, but the tools used to measure language are roughly the same.

\paragraph{Data}
First, we start with a dataset of YouTube and BitChute channels that were linked to in U.S. election fraud discussions on Twitter, compiled in \citet{childs2022characterizing}. The authors combined two publicly-available datasets: the VoterFraud2020 dataset \cite{abilov2021voterfraud2020} and the MeLa-BitChute dataset \cite{trujillo2022mela} to find these channels. From this mapping, the authors produced 13K unique videos from 5084 YouTube channels and 342 BitChute channels that were linked to Tweets discussing U.S. election fraud. The authors found that 36 of the producers (out of the total 5426 channels) had both YouTube and BitChute channels. This matching was done using Levenshtein distance between channel names. 

Using this set of matching channels, we manually confirm that each channel is a legitimate match and filter out channels that do not have more than 20 videos produced during the whole time frame, leaving us with 27 matched channels. We then collect all YouTube and BitChute videos produced between October 2020 and October 2021 by each channel. The dataset contains \textbf{18,815 videos} from \textbf{27 matched channels}. From YouTube, there are 5,610 videos, while on BitChute there are 13,205 videos. To the best of our knowledge, nearly all videos produced by the channels during the time frame are present in the dataset. Hence, as a whole, these channels produced more videos on BitChute during the time frame. We start with this seed dataset to operationalize the task of finding matched channels, as finding all matched channels across YouTube and BitChute is infeasible given the size of YouTube. 

\paragraph{Analysis}
Using this dataset, we analyze the differences between each matched content producer's YouTube channel and BitChute channel using LIWC-22 and Perspective API. Specifically, we compute language features on each channel's video titles across the two platforms. Using these features, we compare differences in video title composition across the platforms using Tukey’s honestly significant difference (HSD) test. Tukey’s HSD test is a method of testing for significant difference between the means of a set of samples. The test uses the difference of the samples to determine which pairs of samples in the presented sample set are significantly different from one another. The null hypothesis is that samples share the same mean, as such, a significant p-value indicates that each of a pair of samples is likely not drawn from the same distribution.

For more information on LIWC-22 please see the write up from \citet{boyd2022development}. For more information on the Perspective API, please see: \url{https://www.perspectiveapi.com}.

\begin{table}[ht]
    \centering
    \small
    \begin{tabular}{c|c|c}
    \toprule
        \textbf{Channel} & \textbf{Created YT} & \textbf{Created BC} \\\midrule
        Styxhexenhammer666 & 2008 & 2017\\
        Resistance Chicks Church  & 2013 & 2018\\
        Martin Brodel  & 2011 & 2019\\
        The Duran  & 2016 & 2018\\
        Timcast  & 2010 & 2017\\
        Tim Pool  & 2011 & 2018\\
        Timcast IRL  & 2016 & 2019\\
        MR. OBVIOUS  & 2017 & 2018\\
        Akkad Daily  & 2019 & 2019\\
        Memology 101 & 2014 & 2018\\
        Mark Dice  & 2007 & 2017\\
        StevenCrowder & 2006 & 2017\\
        Vicious Alien Klown World  & 2021 & 2019\\
        Zoon Politikon  & 2013 & 2017\\
        Dan Bongino & 2013 & 2020\\
        Project Veritas  & 2008 & 2018\\
        Vaccine Wars  & 2020 & 2020\\
        U2Bheavenbound Warrior  & 2013 & 2018\\
        3D to 5D Consciousness  & 2017 & 2020\\
        Misandry Today  & 2016 & 2017\\
        Rebel News & 2015 & 2018\\
        Raw Zone Political Avenger & 2021 & 2017\\
        MatthewHunter1776  & 2020 & 2020\\
        Seb Menard  & 2006 & 2020\\
        The Jordan Report  & 2009 & 2020\\
        Memory Hole Blog  & 2010 & 2019\\
        Just the News  & 2020 & 2020\\\bottomrule
    \end{tabular}
    \caption{The year each channel created their YouTube and BitChute channel. Note, Dan Bongino and Vaccine Wars have been banned from YouTube since our data collection.}
    \label{tab:channels}
\end{table}

\section{Results}

\paragraph{\textbf{Channel Descriptions}} To add some context to the set of content producers we are examining, we provide some metadata of the channels. These channels ranged from 100 to 1,460,000 subscribers on YouTube and 166 to 144,000 subscribers on BitChute. Several of these channels are by well-known content producers in the alt-right space. Given our sampling method, the channels mostly produced content about politics, although several of the channels also produced content about religion and various health-related conspiracy theories. 

Averaged together, these 27 channels used more negative tone (e.g. bad, wrong, hate), more conflict words (e.g. fight, kill, attack), more power words (e.g. own, order, allow, power), more moral words (e.g. wrong, honor, deserve, judge), and more political words (e.g. united states, govern, congress, senate) in the titles of their videos on BitChute than the average BitChute video (see \cite{horne2022psycho} for platform-level averages). All four categories are more than the average content on Facebook, Twitter, and Reddit. Similarly, together these channels used more negative tone and more political words in the titles of their videos on YouTube than the average BitChute video. However, on average their YouTube video titles use less conflict words, less power words, and less moral words than the average BitChute video. These aggregated differences are partially due to the topical focus of these channels (channels that cover U.S. politics). These differences also indicate that co-active YouTube channels and BitChute channels do not necessarily produce the exact same content across platforms. 

\paragraph{\textbf{RQ1: Production across platforms}}
In Table \ref{tab:channels}, we show the 27 matched channels and the years in which they created their YouTube channel and their BitChute channel. Somewhat surprisingly, the YouTube channel is not always created before the BitChute channel. Of the 27 matched channels, two channels created their BitChute channel before creating their YouTube channel, and three channels created both channels at approximately the same time. However, the majority (22 of 27) created their YouTube channels prior to their BitChute channels, with the maximum distance between creations of 14 years and minimum distances between creations of 1 year.

In Figure \ref{tbl:prod}, we show plots of the number of videos published by each channel on each platform over time (October 2020 to October 2021). At a high level, the patterns of production across platforms is very inconsistent across matched channels, with some channels creating much more content on YouTube than BitChute and vice versa. Some channels exhibit the production pattern we expect based on the narrative that channels are moving away from YouTube to BitChute. For example, three channels produced more YouTube videos during 2020, but transition to producing more BitChute videos in 2021. Yet, seven channels show the opposite behavior, producing more on BitChute early in the time frame and more on YouTube late. Some of these channels seem to abandon one platform or another during the time span: channel \texttt{Raw Zone Political Avenger} stopped producing videos on BitChute in July 2021, channel \texttt{Seb Menard} stopped producing videos on BitChute in December 2020, and channel \texttt{Vaccine Wars} stopped producing videos on YouTube in January 2021 (because they were banned by YouTube). Others picked up one platform during the time span: channel \texttt{Vicious Allen Klown World} did not produce YouTube videos until the middle of 2021 (when their YouTube channel was created) and channel \texttt{Resistance Chicks Church} did not produced YouTube videos until the middle of 2021 (despite their YouTube channel existing since 2013). 

Overall, there is not a single pattern of production across platforms in our data sample. Perhaps most importantly, this lack of pattern demonstrates that today channels on BitChute are not necessarily clones of channels on YouTube, opposite of what evidence in early studies of BitChute demonstrated \cite{trujillo2020bitchute}. Although in a few of our sampled channels, the videos are the same on YouTube and BitChute (for example the channel Styxhexenhammer666).

\begin{table}[ht]
    \centering
    \begin{tabular}{c|c|c}
        \textbf{Feature} & \textbf{\# BC channels more} & \textbf{\# YT channels more} \\\midrule
        toxicity & 1 of 27 & 3 of 27  \\
        Tone & 11 of 27 & 5 of 27 \\
        affiliation & 1 of 27 & 10 of 27 \\
        achieve & 9 of 27 & 3 of 27 \\
        power & 13 of 27 & 2 of 27 \\
        certitude & 7 of 27 & 2 of 27 \\
        cause & 11 of 27 & 3 of 27 \\
        tone\_neg & 4 of 27 & 12 of 27 \\
        emo\_anger & 6 of 27& 0 of 27 \\
        prosocial & 0 of 27 & 11 of 27\\
        polite & 3 of 27& 1 of 27 \\
        conflict & 11 of 27& 0 of 27 \\
        moral & \textbf{14 of 27}& 0 of 27 \\
        politic & \textbf{16 of 27}& 1 of 27\\
        ethnicity & 8 of 27& 1 of 27 \\
        relig & 7 of 27& 1 of 27 \\
        sexual & 4 of 27& 0 of 27\\
        death & 9 of 27& 0 of 27 \\
        need &7 of 27& 1 of 27\\
        we & 9 of 27& 3 of 27\\
        male & 8 of 27& 6 of 27\\
        female & 4 of 27& 1 of 27\\\bottomrule
    \end{tabular}
    \caption{The number of channels in the dataset that showed significantly more use of a feature on BitChute (BC) or YouTube (YT) in video titles, where significance is determined by Tukey’s HSD. We note in bold the features where the majority of channels ($> 50\%$) are significantly higher.}
    \label{tab:majority}
\end{table}

\paragraph{\textbf{RQ2: Toxicity and language across platforms}}
Next, we examine the toxicity and language use in video titles across platforms using the Perspective API and LIWC-22. In Table \ref{tab:majority}, we show the number of channels that showed significant differences in language and toxicity use across 22 features. In Figures \ref{featfigs} and \ref{featfigs2}, we show a selection of channel-by-channel feature differences.

As shown in Table \ref{tab:majority}, the majority of content producers used significantly more political words (e.g. united states, govern, congress, senate) and moral words (e.g. wrong, honor, deserve, judge) in their video titles on BitChute compared to their video titles on YouTube. Other features that were often used more in titles on BitChute, but not by the majority of content producers, included power words (e.g. own, order, allow, power), casual words (e.g. how, because, make, why), and conflict words (e.g. fight, kill, attack), used more on BitChute by 48\%, 40\%, and 40\% of the content producers, respectively. On the other hand, content producers often used higher levels of negative tone (e.g. bad, wrong, too much, hate) and prosocial language (e.g. care, help, thank, please) in their video titles on YouTube, used more on YouTube by 44\% and 40\% of content producers, respectively.

Notably, while there are some differences in the language use in video titles across platforms (particularly in the use of moral and political language), the majority of channels do not use significantly different language across the platforms, despite often producing different videos across the platforms during the same time span. Further, unexpectedly, the toxicity of video titles does not change significantly across the platforms. Only 1 channel used more toxicity in their video titles on BitChute than on YouTube, while 3 channels used more toxicity in their video titles on YouTube than BitChute. In both cases, the effect sizes are very small. 

Of course, these trends are not one-size-fits-all, with some clear outliers shown in the channel-by-channel differences. For example, when examining the channel \texttt{The Jordan Report}, our language features showed a significantly higher use of ethnicity words in their BitChute titles. When qualitatively examining videos from \texttt{The Jordan Report} on both platforms, we saw consistent themes of racism and antisemitism across both platforms, but confirm that there was a shift in ethnicity word usage on BitChute. We found the opposite to be true about the channel \texttt{U2Bheavenbound Warrior}, which used significantly more ethnicity words in their YouTube titles. Interestingly, when qualitatively examining the channel on both platforms, we found that this word usage difference was due to each platform having a different topical focus during the time-frame. Namely, \texttt{U2Bheavenbound Warrior}'s YouTube channel produced videos on global "end times" conspiracies in the US, UK, Russia, Israel, and China. While, during the same time frame on BitChute, videos were produced about the 2020 US Elections and COVID-19 conspiracy theories. This different topical production across the platforms may have been due to differences in what content was moderated on YouTube, as we know YouTube increased moderation around election and COVID-19 conspiracy theories during that time.

\section{Conclusion}
In this short paper, we provide the first exploration of co-active content creators on YouTube and BitChute. We find evidence that co-active channels produce different content across the platforms but that content tends to not significantly differ in language characteristics. In some cases where language did significantly differ across the platforms, the content was about completely different topics. In other cases, these changes were due to platform production changes across time (event coverage). In rare cases, video titles of the same video were being changed across the platforms. However, no single pattern was consistent across the 27 matched channels. With a broader sample of co-active channels, future work can better categorize the reasons and behavior across the platforms.

This inconsistency points to a broader point: production on and movement to alt-tech social media platforms does not fit a single narrative. While yes, it is still clear that some content producers move to alt-tech platforms due to being banned from from mainstream platforms, this narrative does not fit all content producers. In this case of this study, we show that content producers may change topics completely across platforms, perhaps producing content for a different audience or to avoid moderation measures on specific content. We also found that some producers move the opposite direction (from alt-tech to mainstream). Future work should better explore this point. While video content may not always differ, the audiences consuming the content does. The audience change across platforms is likely a factor in producers decisions to be on one platform or both platforms. 

\section{Limitations}
This study is not without limitations and should be thought of as a preliminary exploration of co-activity. One limitation is that we focused on operationalizing the matching task rather than extracting the broadest possible sample of matching channels across the platforms. We used a strict channel name match and performed this matching task on a topic and time specific dataset (channels linked to in the VoterFraud2020 dataset). Hence, it is likely that there are many more coexisting channels that were not examined in our analysis and those channels may produce different types of content (in particular content not related to politics). However, this small number of matched channels made manual analysis at the channel level possible, giving us assurances that the channels were legitimately coexisting channels.

Second, given that our data only captures content production during an overlapping period across the platforms, rather than before and after a producer begin creating content on one platform or another, we cannot establish causal relationships or spillover effects of being co-active. In future work, comparing these trends to a baseline set of disjoint content producers from each platform may provide more robust insights.

\section{Ethics Statement}
While we considered removing specific channel names from our study, this removal would reduce our study's provenance and context. Further, the content producers examined in this study post videos for public consumption, hence they should not have the expectation of anonymity.

\begin{table*}[ht]
\large
\begin{tabular}{lllll}
& \includegraphics[width=.22\linewidth,valign=m]{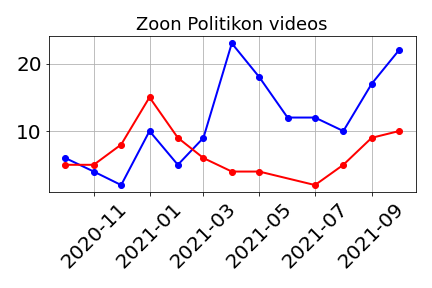} & \includegraphics[width=.22\linewidth,valign=m]{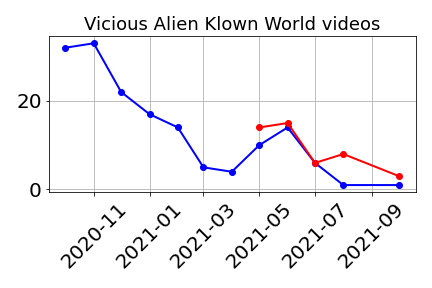} & \includegraphics[width=.22\linewidth,valign=m]{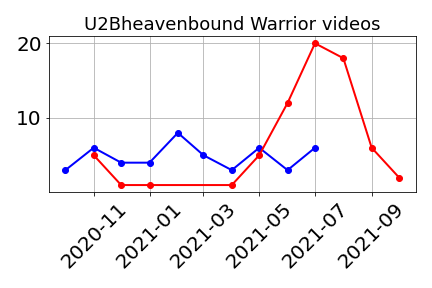}&
\includegraphics[width=.22\linewidth,valign=m]{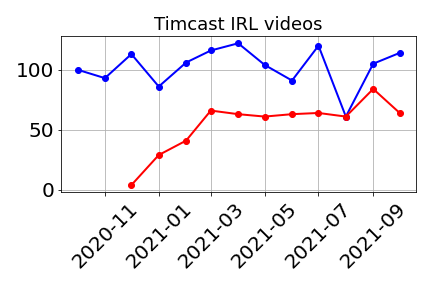}\\
\parbox[t]{3mm}{\multirow{9}{*}{\rotatebox[origin=c]{90}{\textbf{Number of Videos}}}} & \includegraphics[width=.22\linewidth,valign=m]{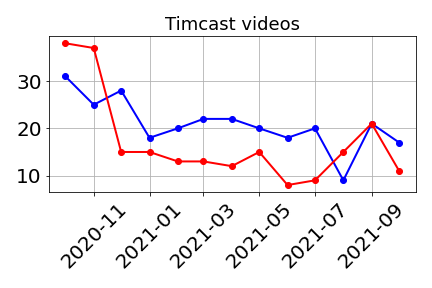} & \includegraphics[width=.22\linewidth,valign=m]{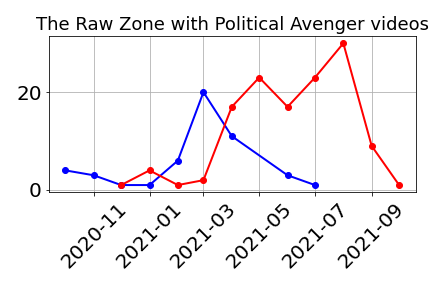} & \includegraphics[width=.22\linewidth,valign=m]{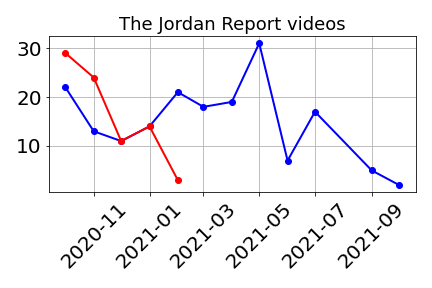}&
\includegraphics[width=.22\linewidth,valign=m]{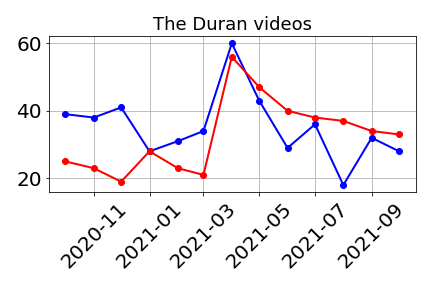}\\
& \includegraphics[width=.22\linewidth,valign=m]{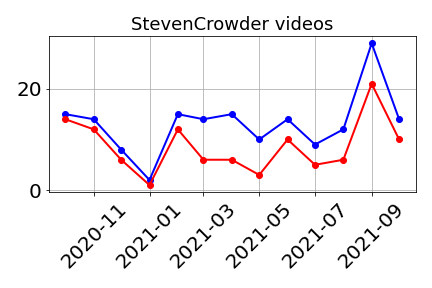} & \includegraphics[width=.22\linewidth,valign=m]{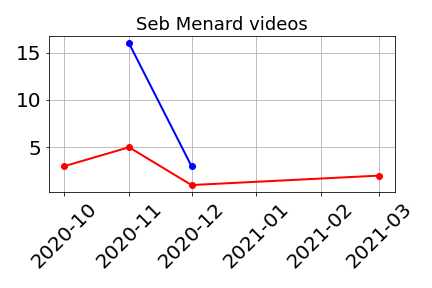} & \includegraphics[width=.22\linewidth,valign=m]{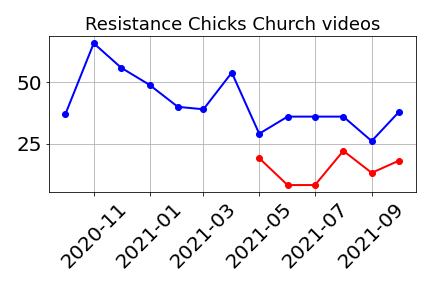}&
\includegraphics[width=.22\linewidth,valign=m]{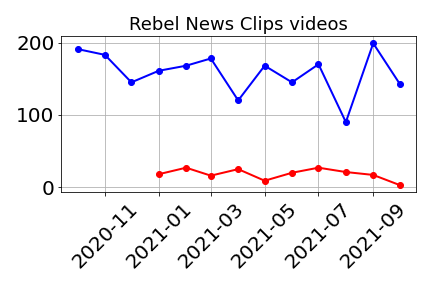}\\
& \includegraphics[width=.22\linewidth,valign=m]{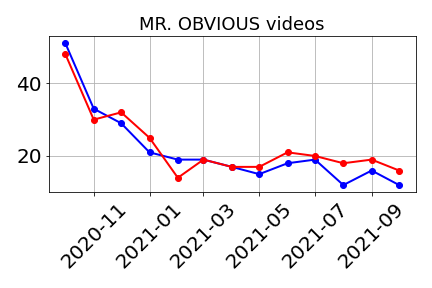} & \includegraphics[width=.22\linewidth,valign=m]{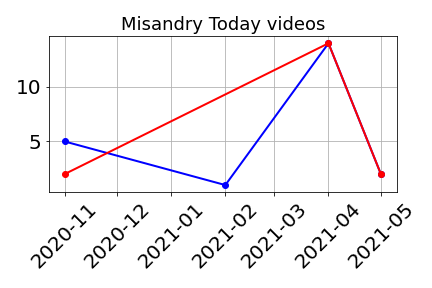} & \includegraphics[width=.22\linewidth,valign=m]{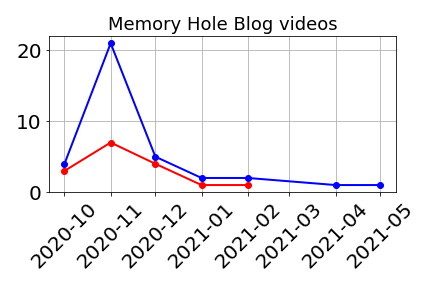}&
\includegraphics[width=.22\linewidth,valign=m]{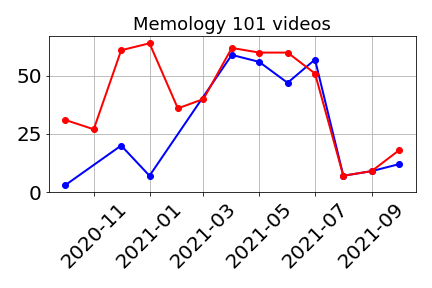}\\
& \includegraphics[width=.22\linewidth,valign=m]{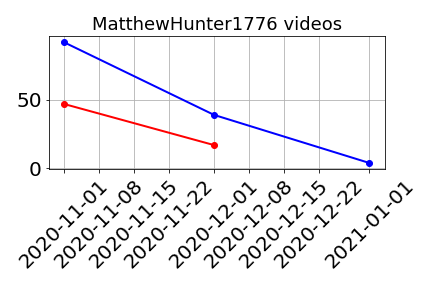} & \includegraphics[width=.22\linewidth,valign=m]{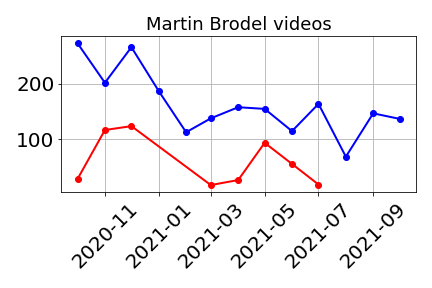} & \includegraphics[width=.22\linewidth,valign=m]{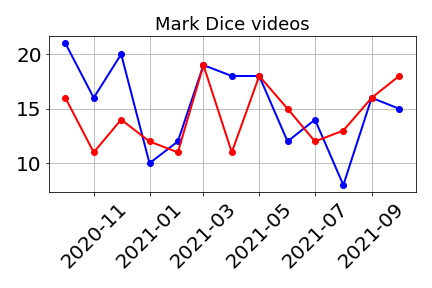}&
\includegraphics[width=.22\linewidth,valign=m]{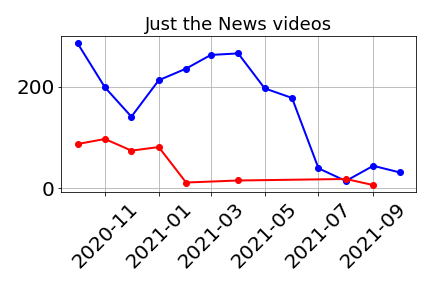}\\
& \includegraphics[width=.22\linewidth,valign=m]{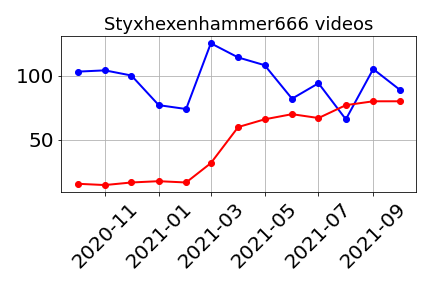} & \includegraphics[width=.22\linewidth,valign=m]{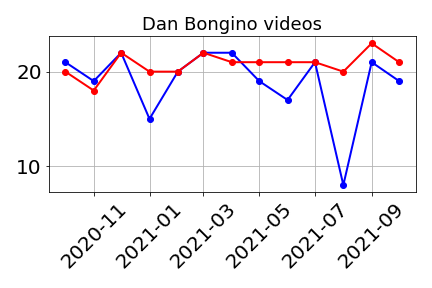} & \includegraphics[width=.22\linewidth,valign=m]{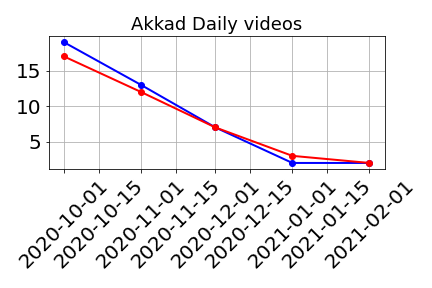}&
\includegraphics[width=.22\linewidth,valign=m]{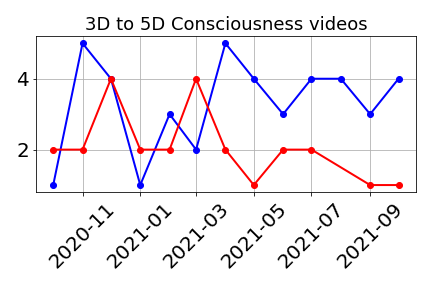}\\
& \includegraphics[width=.22\linewidth,valign=m]{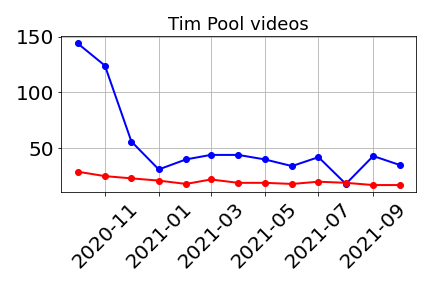} & \includegraphics[width=.22\linewidth,valign=m]{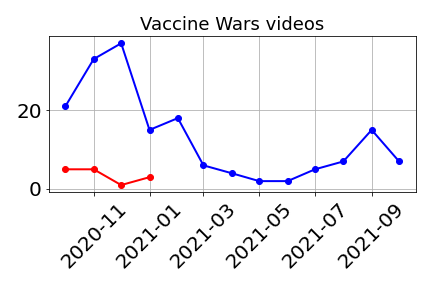} & \includegraphics[width=.22\linewidth,valign=m]{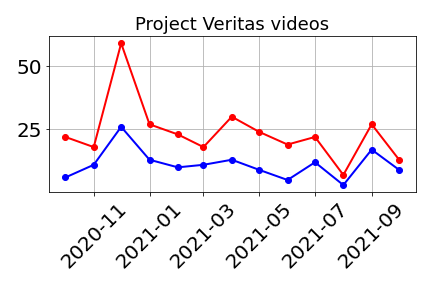} & \includegraphics[width=.15\linewidth,valign=m]{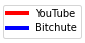}
\\\addlinespace[3mm]
\multicolumn{5}{c}{\textbf{Months (between Oct 2020 and Oct 2021)}} \\
\end{tabular}
\captionof{figure}{Number of videos published over time per channel per platform, where red is YouTube and blue is BitChute. Note the differences in the x-axes, as each channel produces varying number of videos over all.}
\label{tbl:prod}
\end{table*}

\begin{table*}[ht]
\large
\begin{tabular}{cc}
\includegraphics[width=.5\textwidth,valign=m]{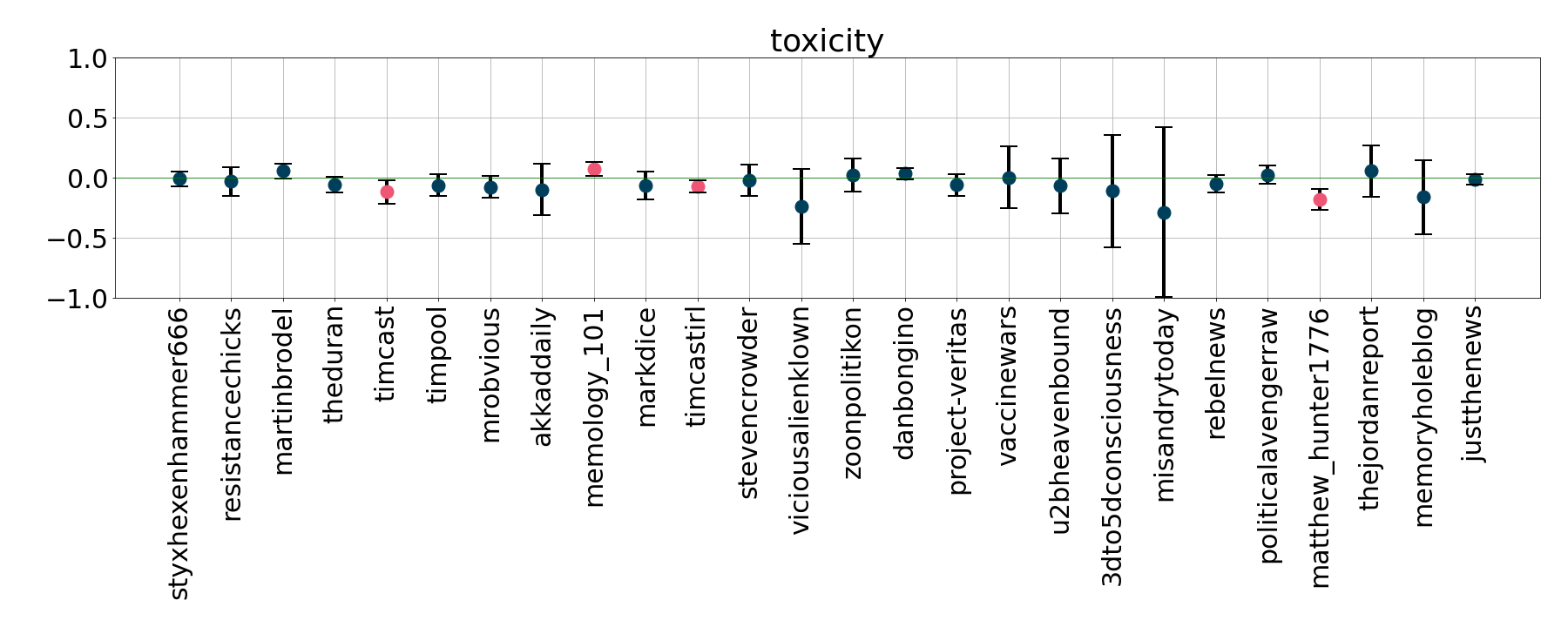} &
\includegraphics[width=.5\textwidth,valign=m]{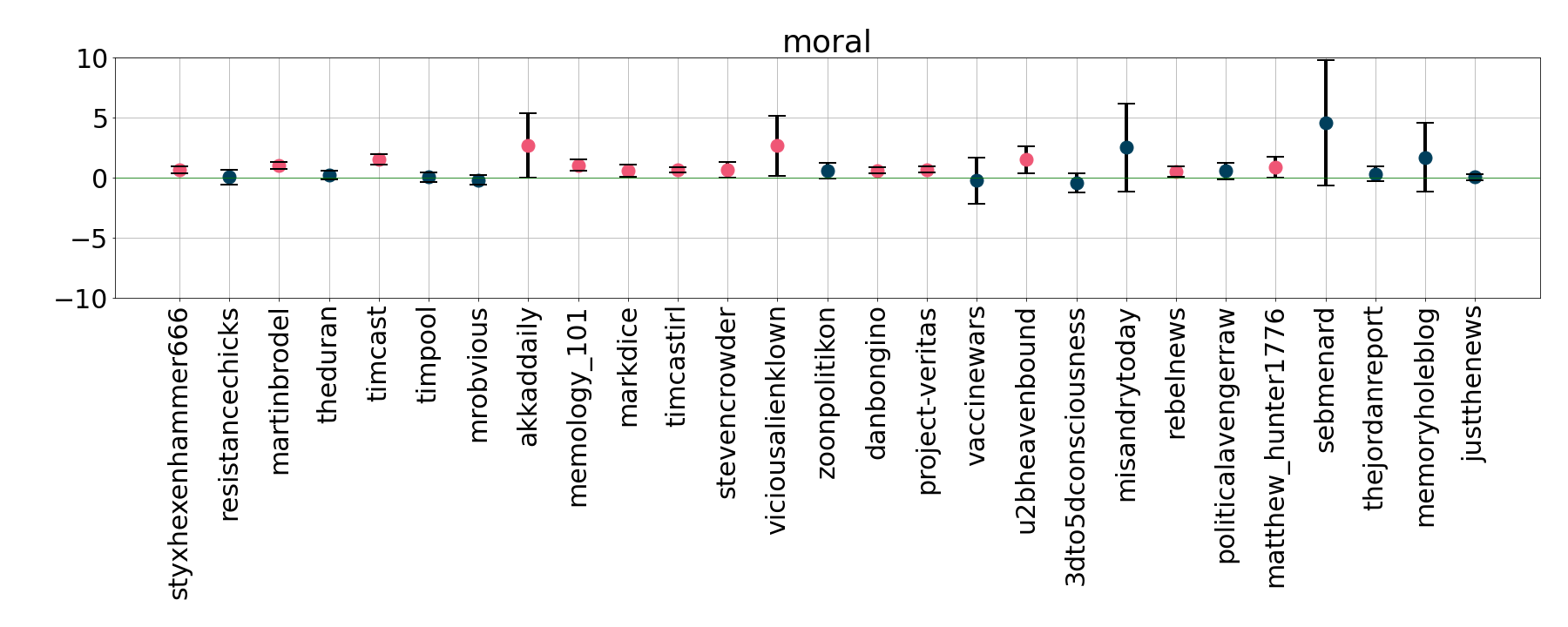}\\
\includegraphics[width=.5\textwidth,valign=m]{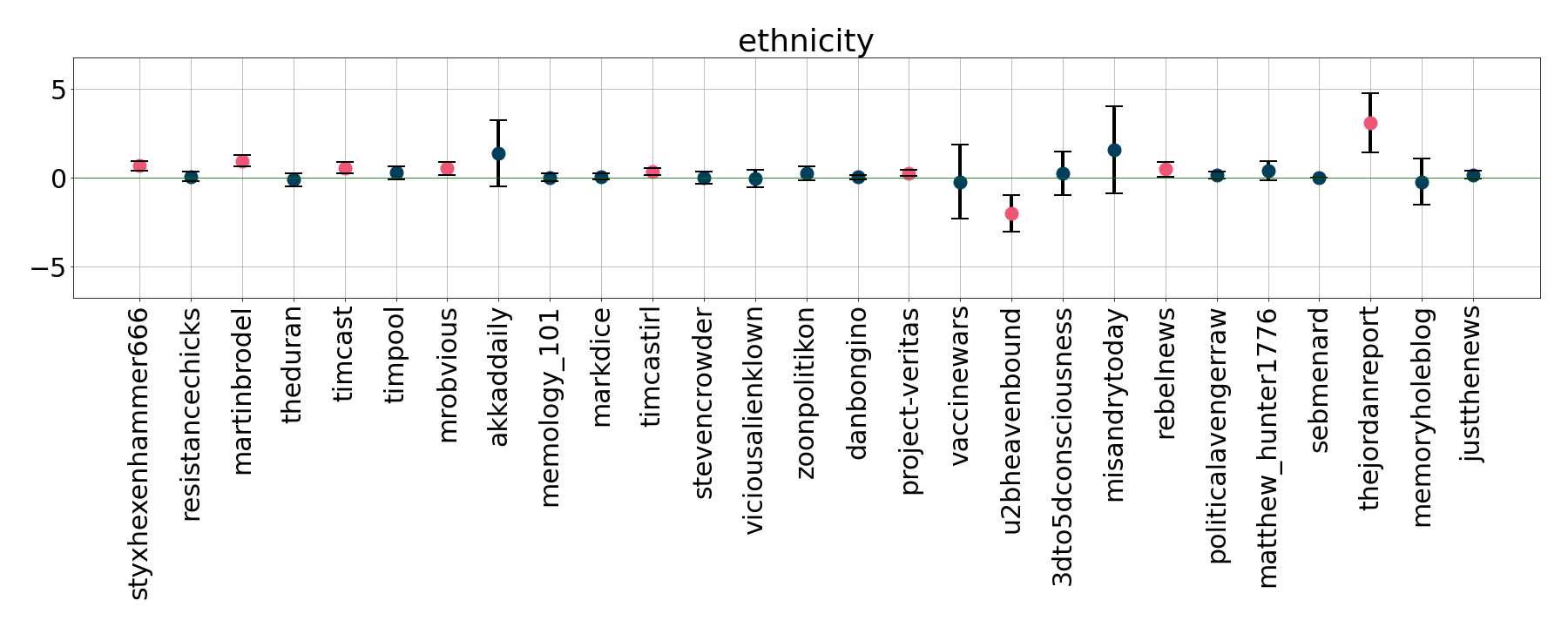} &
\includegraphics[width=.5\textwidth,valign=m]{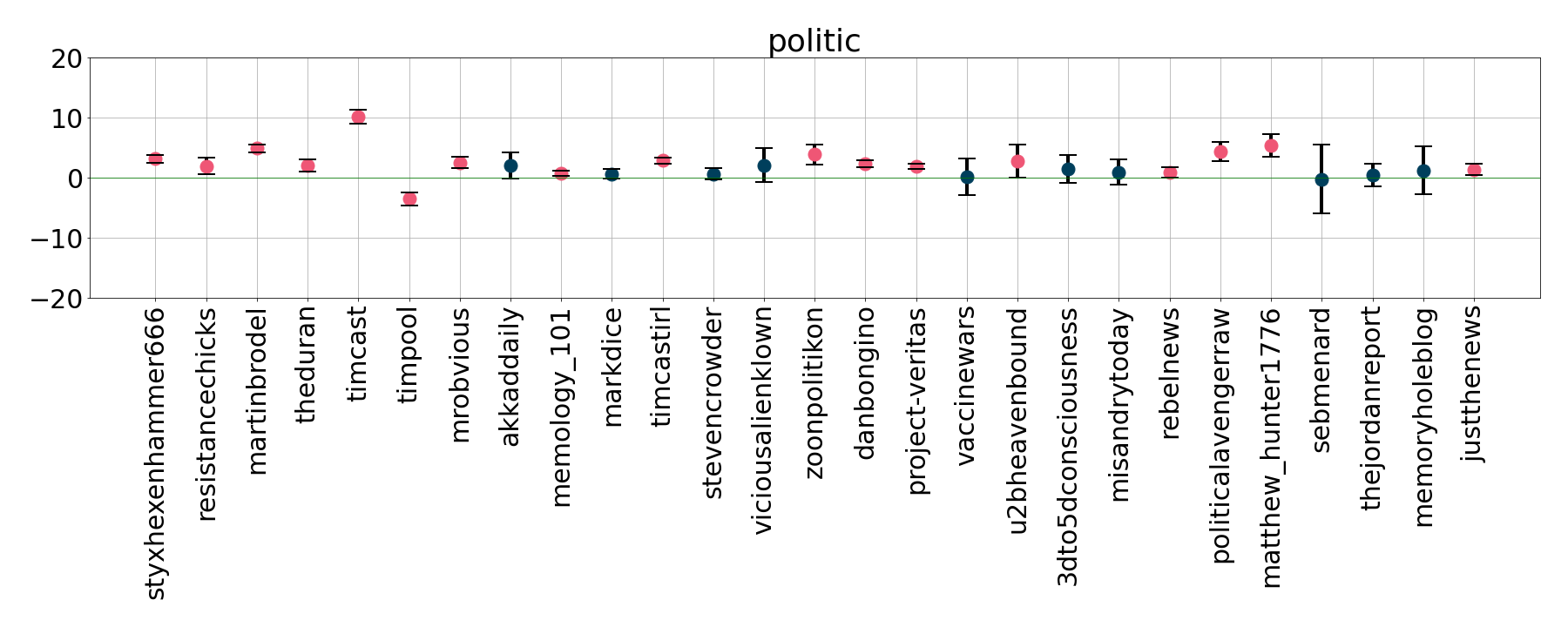}\\
\includegraphics[width=.5\textwidth,valign=m]{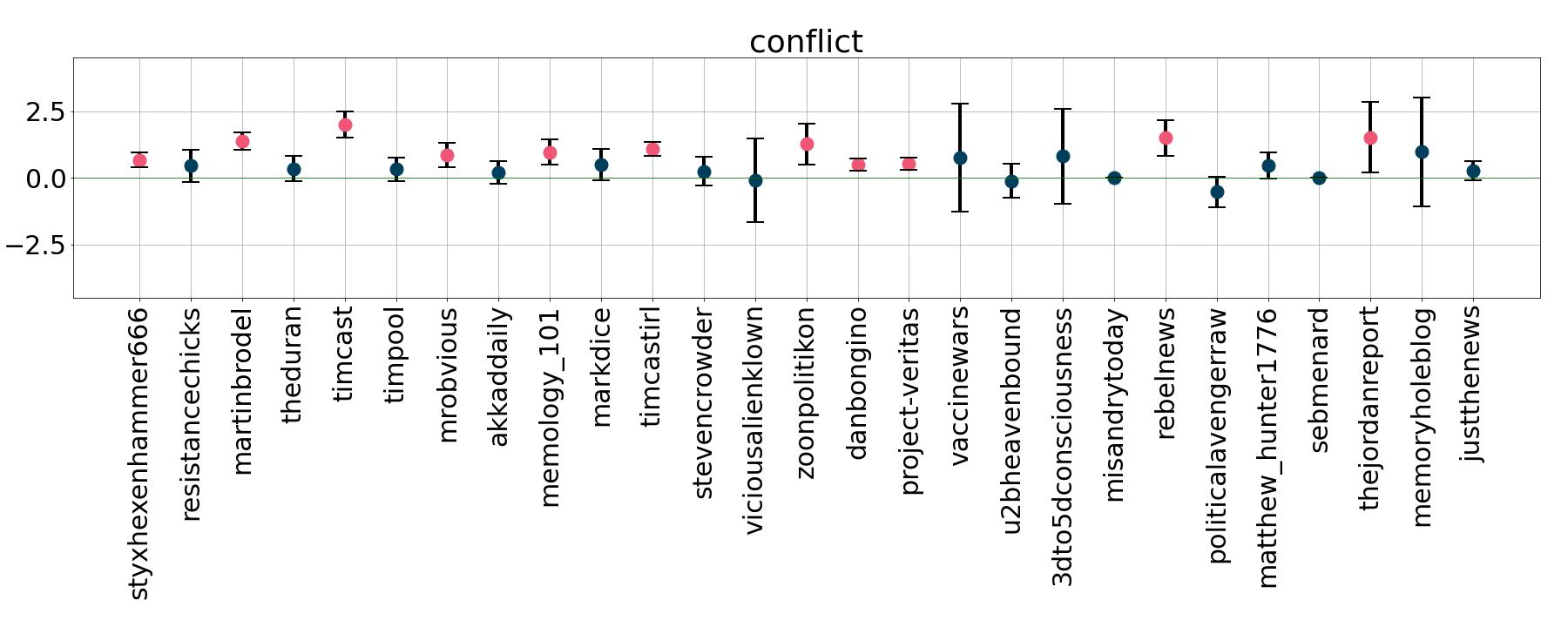}&
\includegraphics[width=.5\textwidth,valign=m]{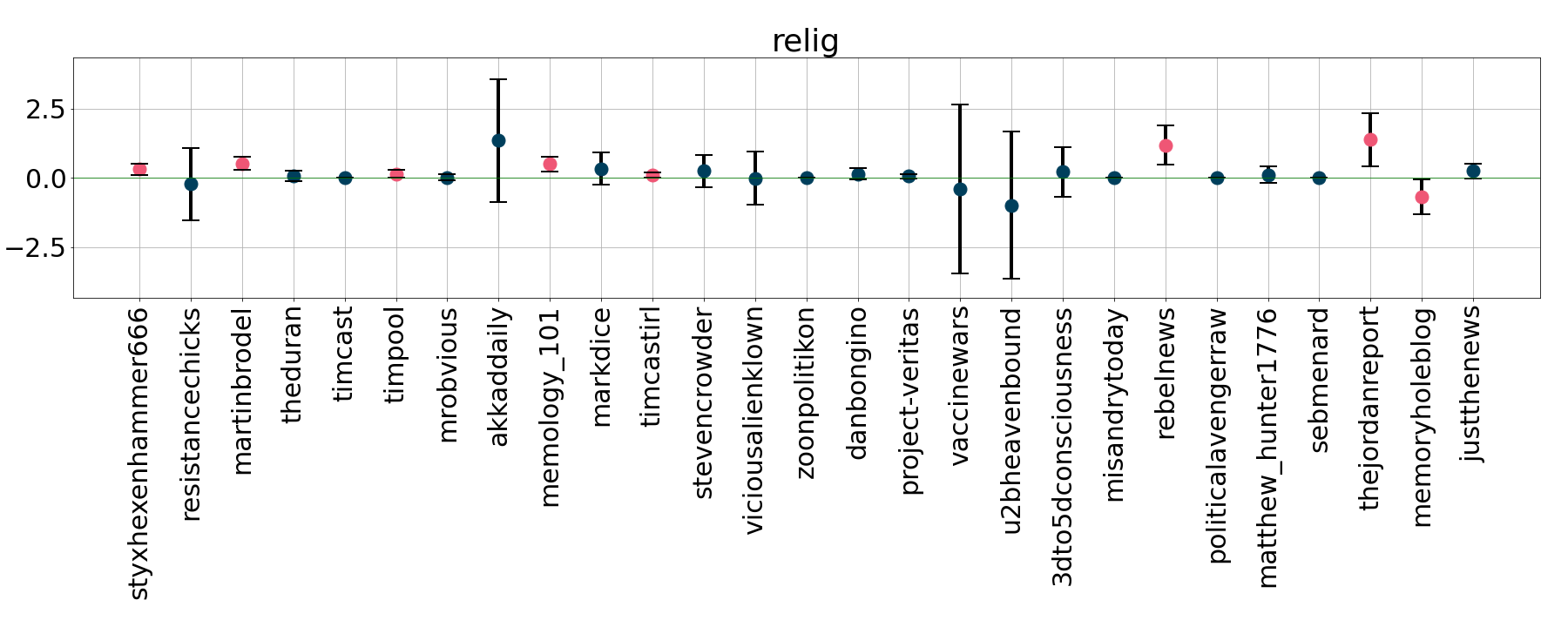}\\
\includegraphics[width=.5\textwidth,valign=m]{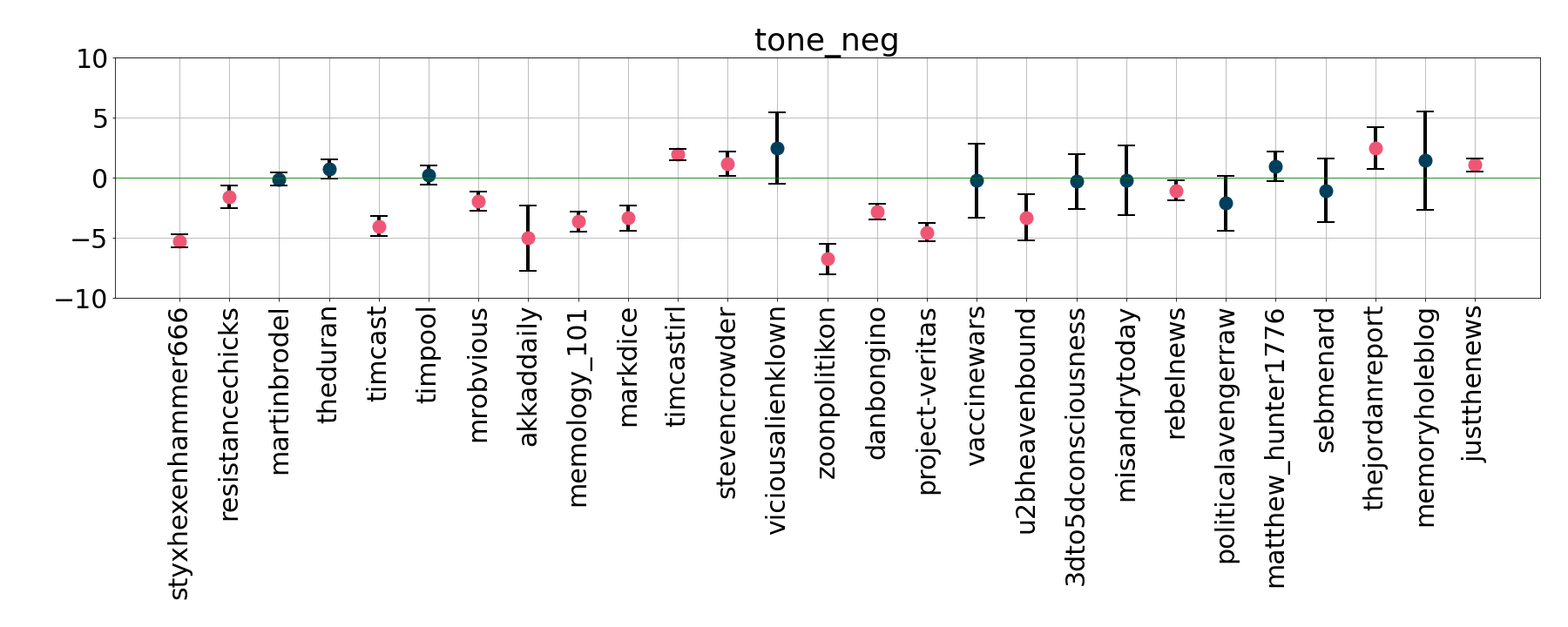}&
\includegraphics[width=.5\textwidth,valign=m]{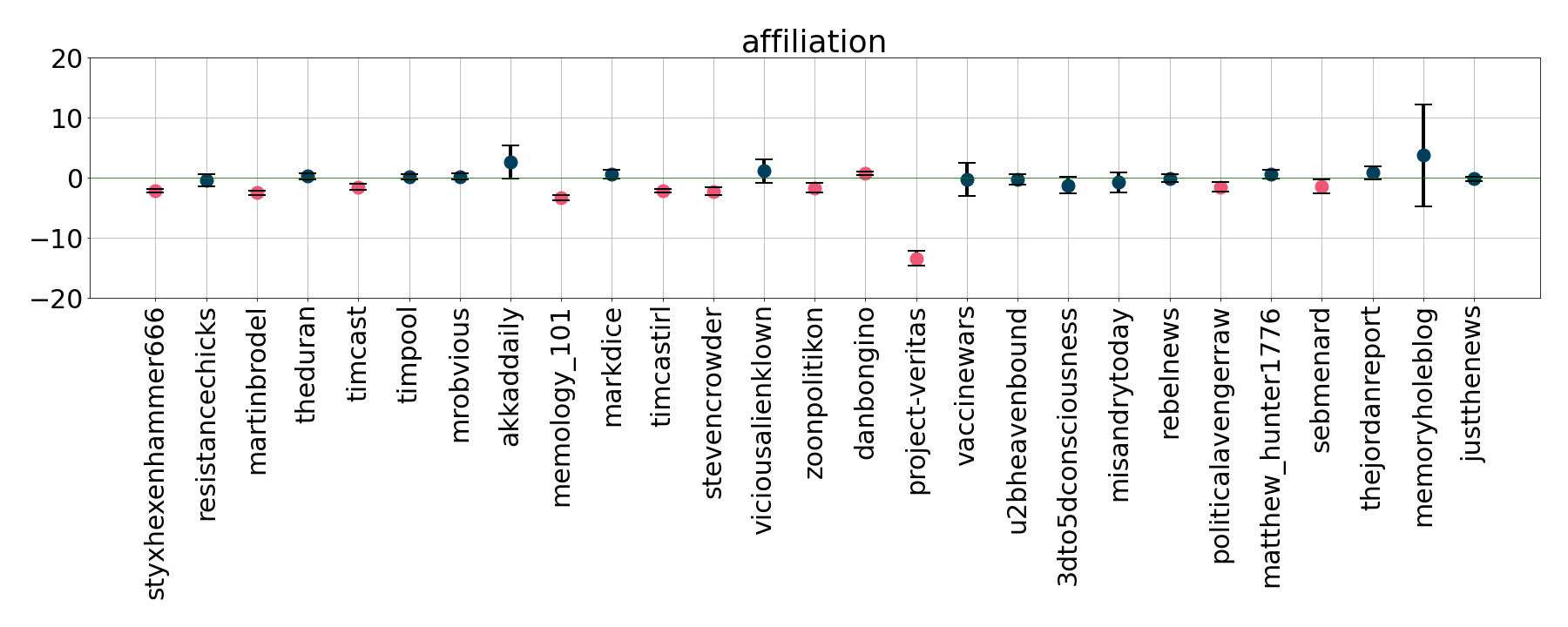}\\
\includegraphics[width=.5\textwidth,valign=m]{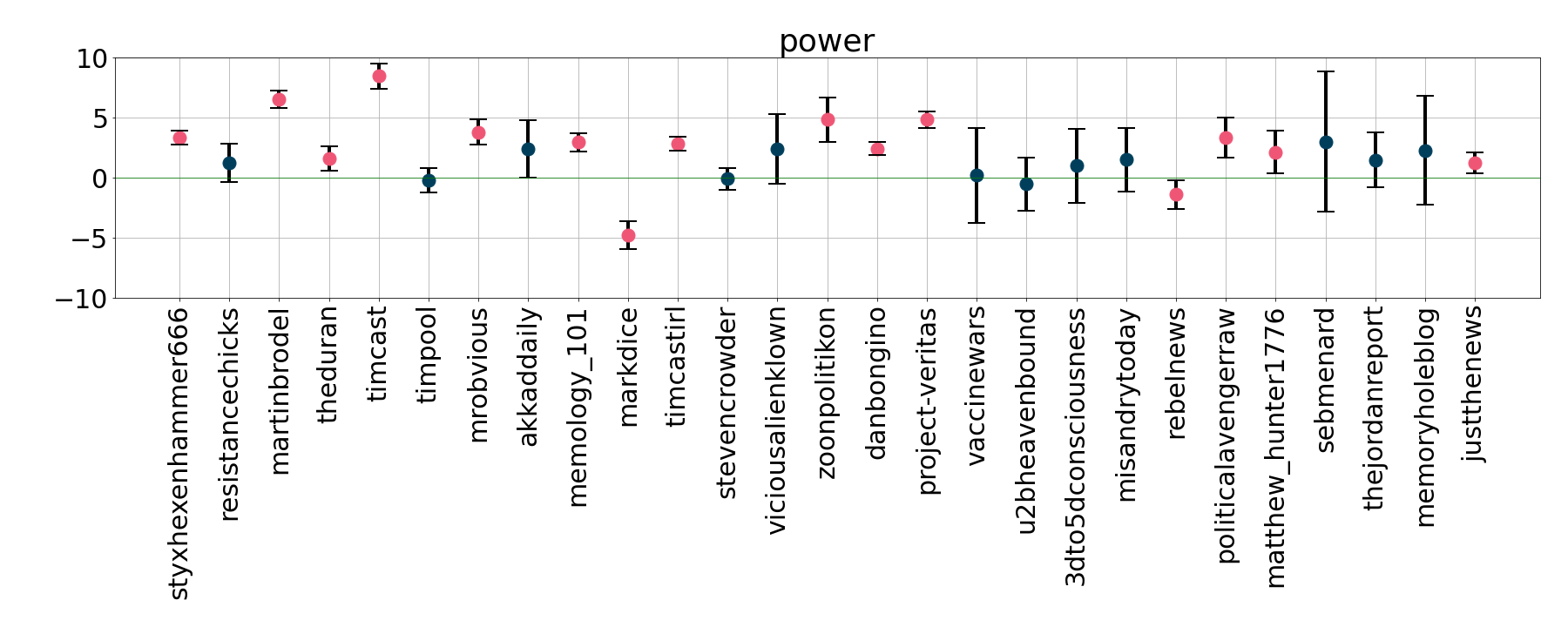}&
\includegraphics[width=.5\textwidth,valign=m]{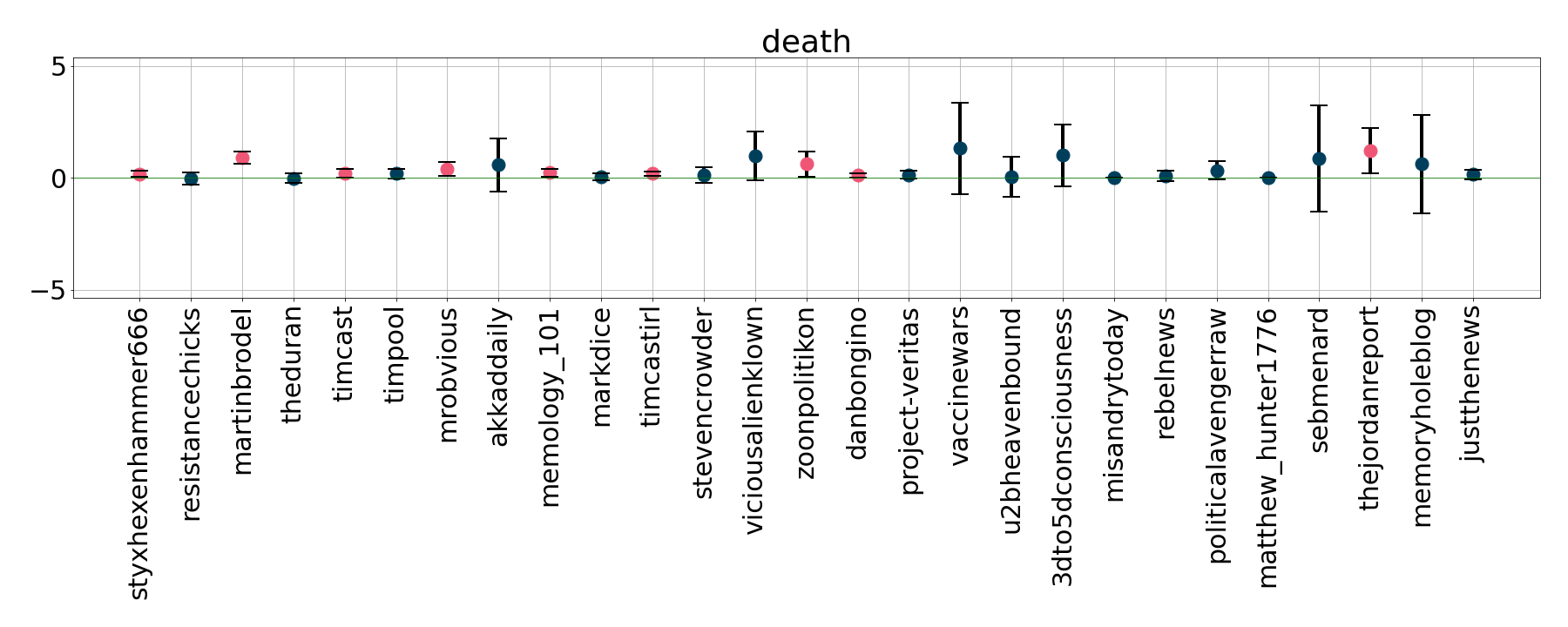}\\
\multicolumn{2}{c}{\includegraphics[width=.11\linewidth, valign=m]{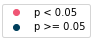}}\\
\end{tabular}
\captionof{figure}{Channel-by-channel differences across platforms of toxicity and selected LIWC features, where negative values indicates the feature is greater on YouTube, positive values indicate the feature is greater on BitChute, and zero indicates no difference across platforms. If those differences are significant according to Tukey’s HSD, they are highlighted in red.}
\label{featfigs}
\end{table*}

\begin{table*}[ht]
\large
\begin{tabular}{cc}
\includegraphics[width=.5\textwidth,valign=m]{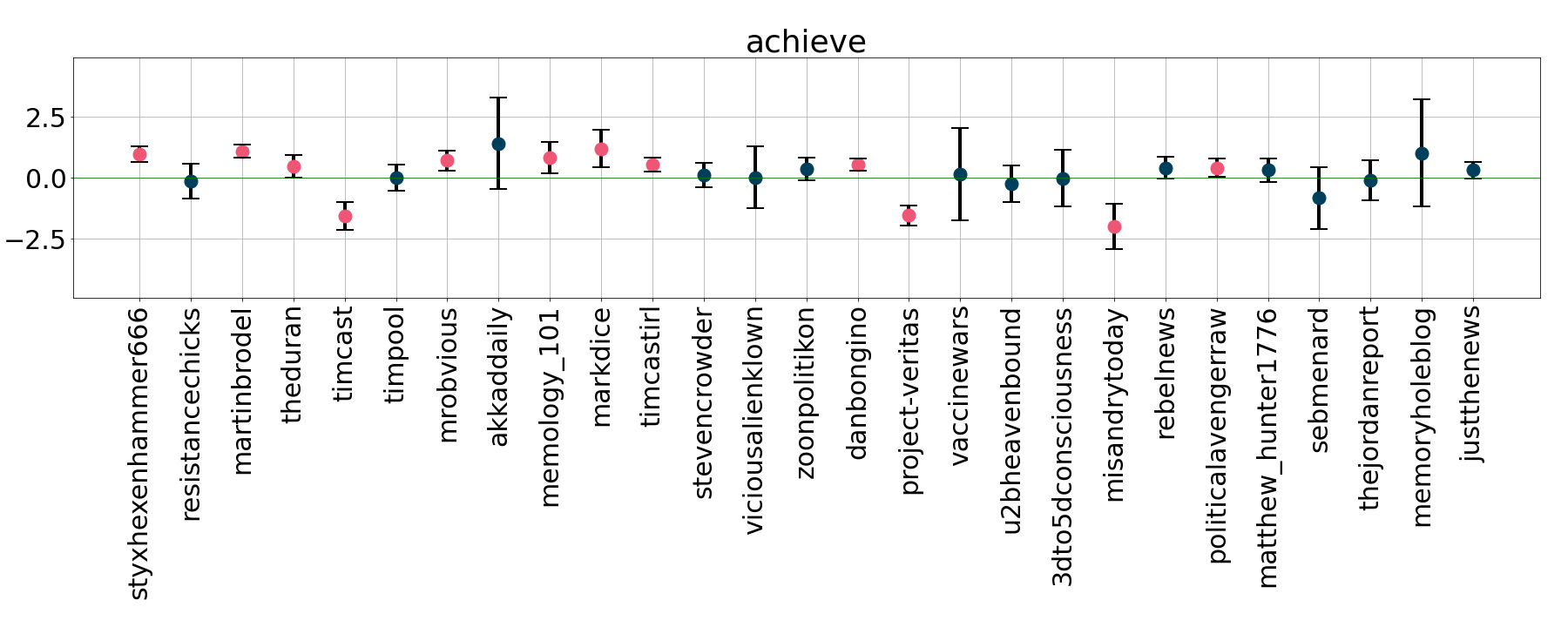} &
\includegraphics[width=.5\textwidth,valign=m]{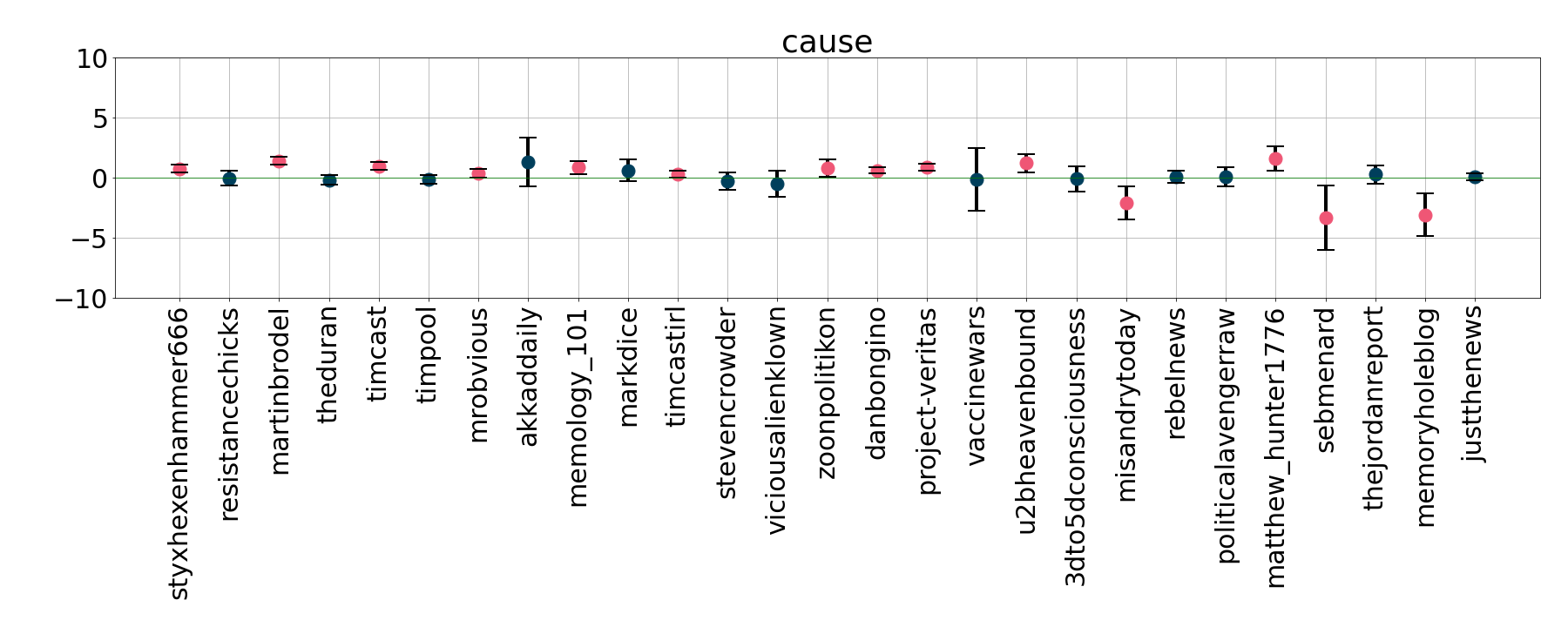}\\
\includegraphics[width=.5\textwidth,valign=m]{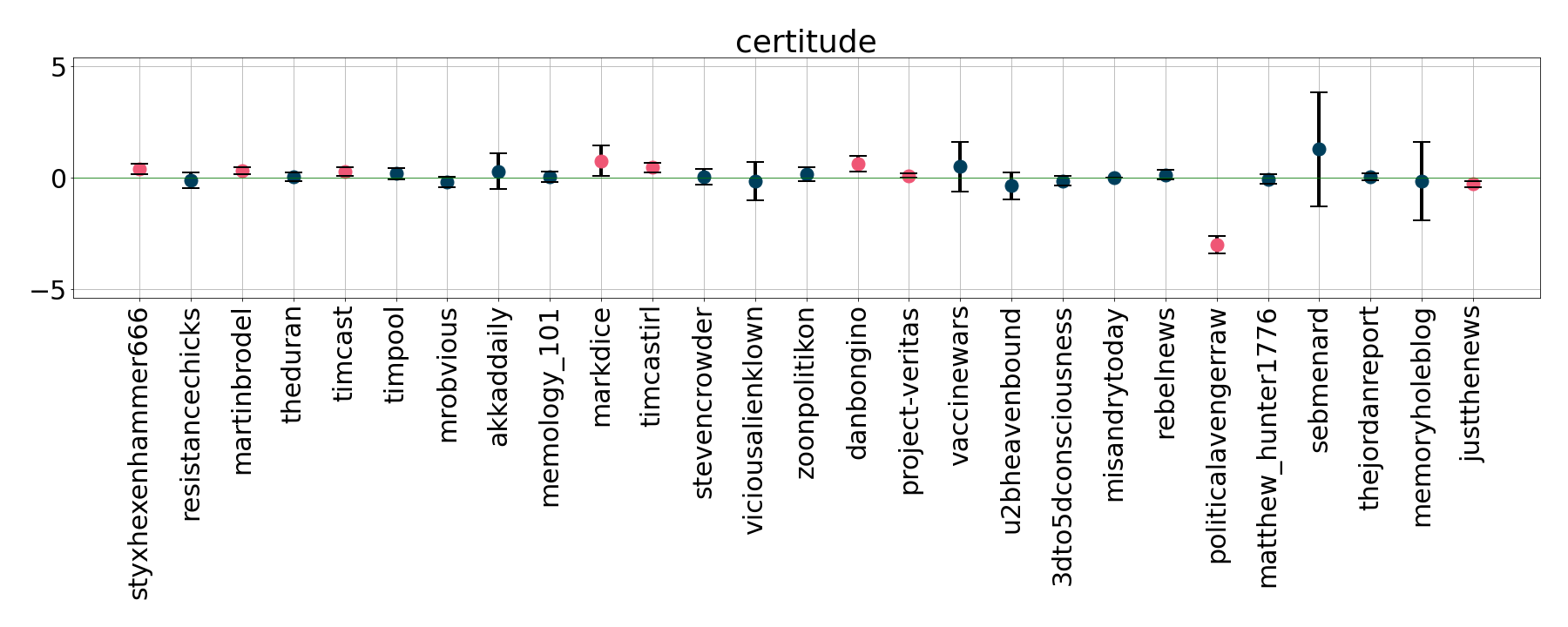} &
\includegraphics[width=.5\textwidth,valign=m]{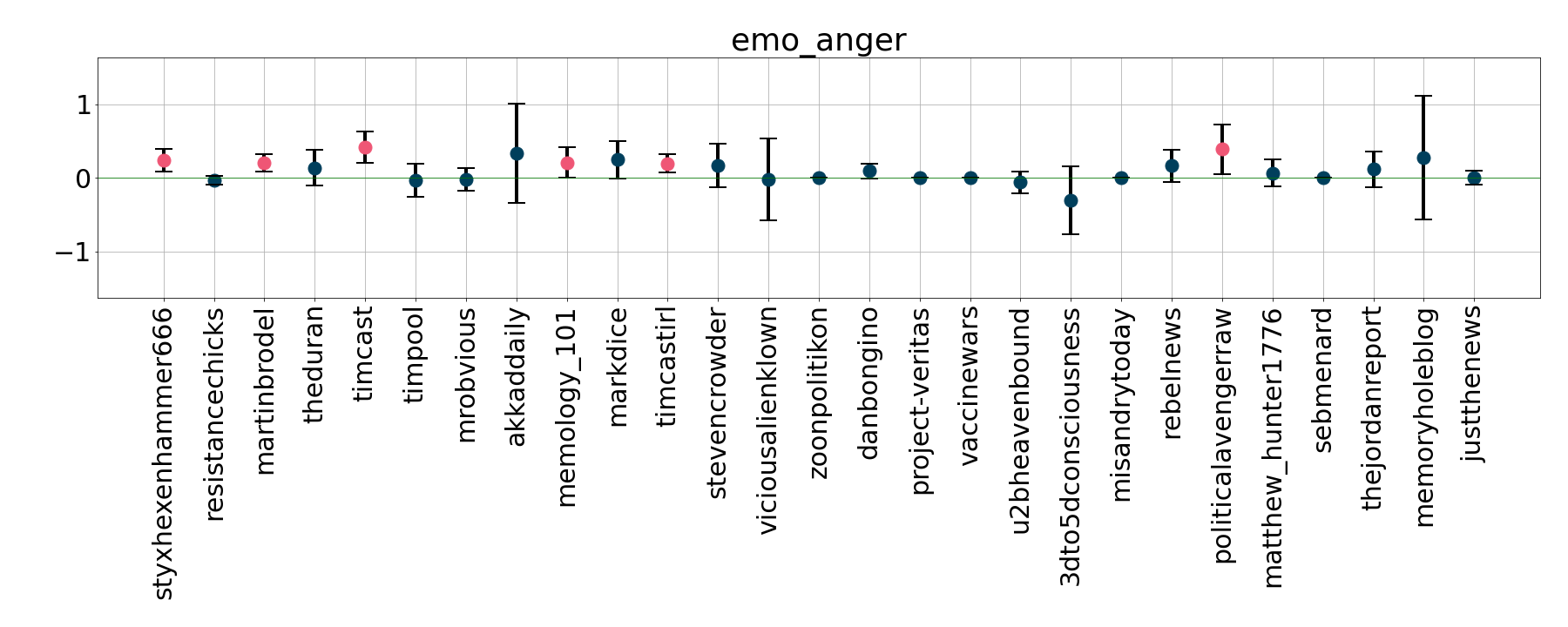}\\
\includegraphics[width=.5\textwidth,valign=m]{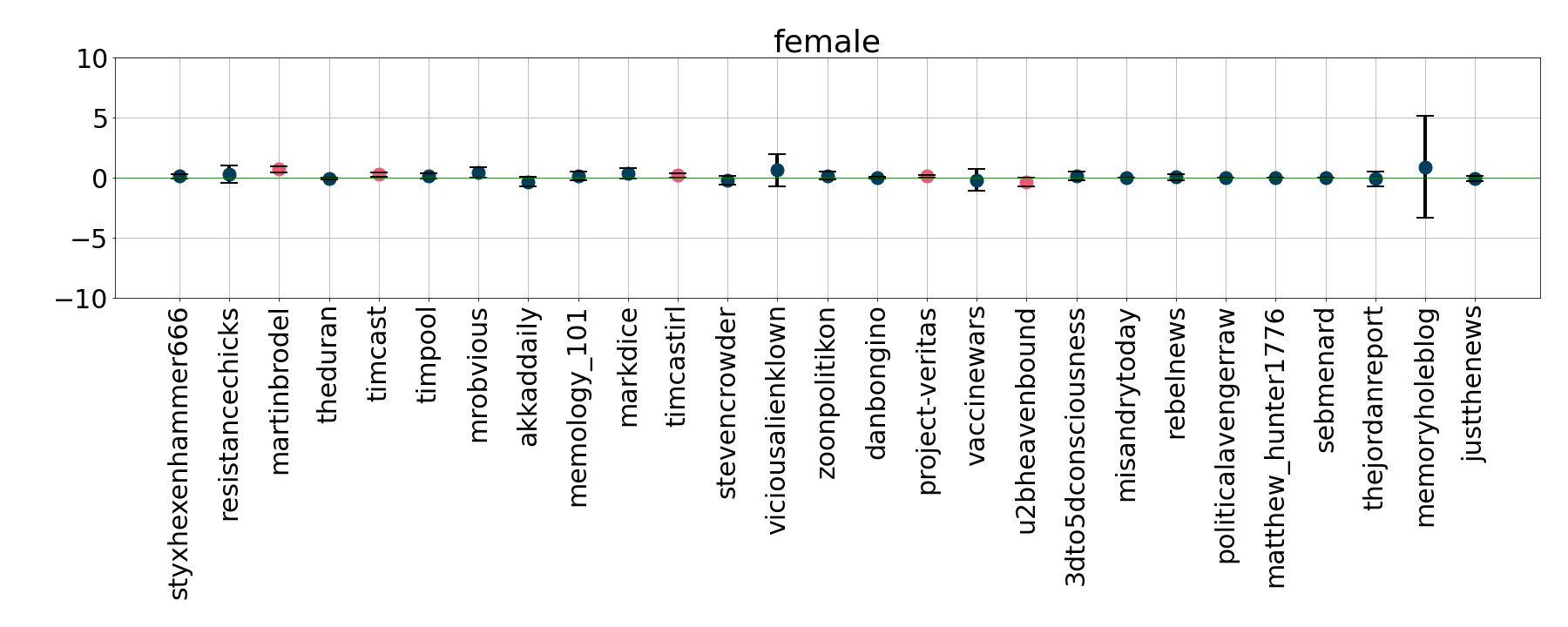}&
\includegraphics[width=.5\textwidth,valign=m]{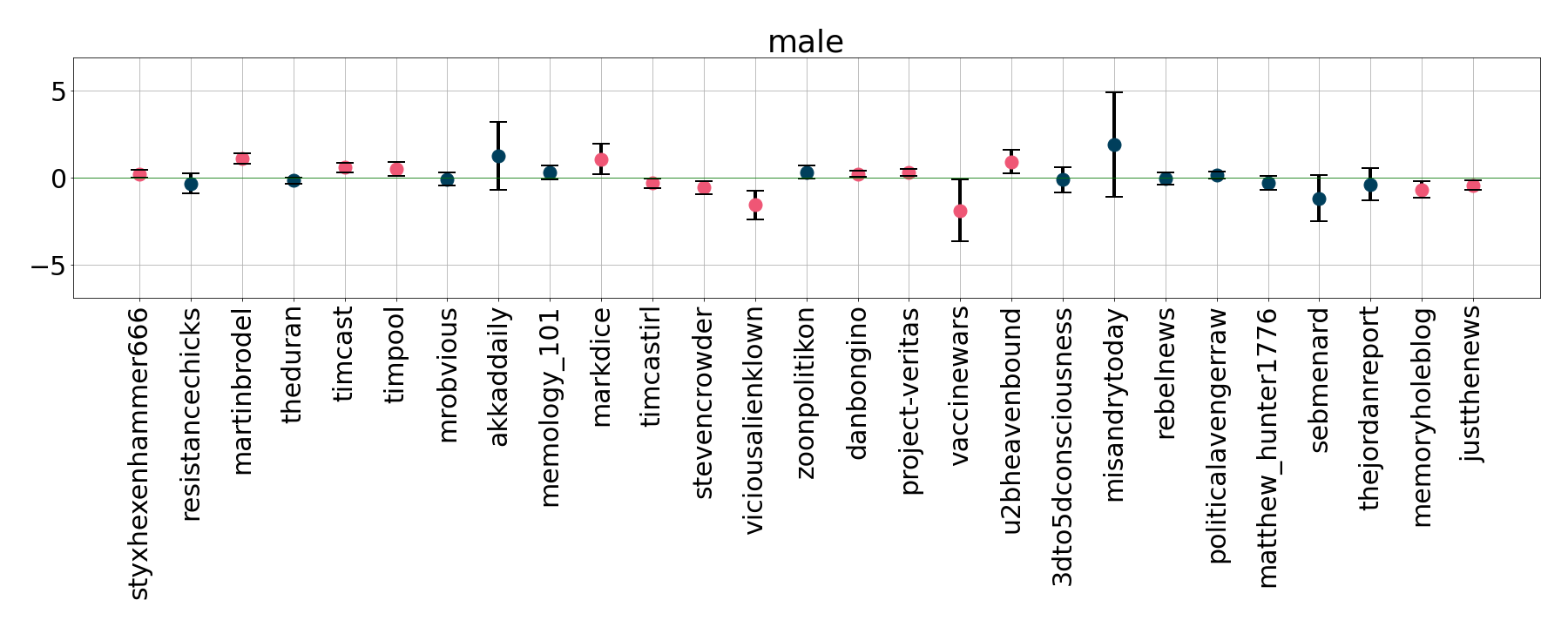}\\
\includegraphics[width=.5\textwidth,valign=m]{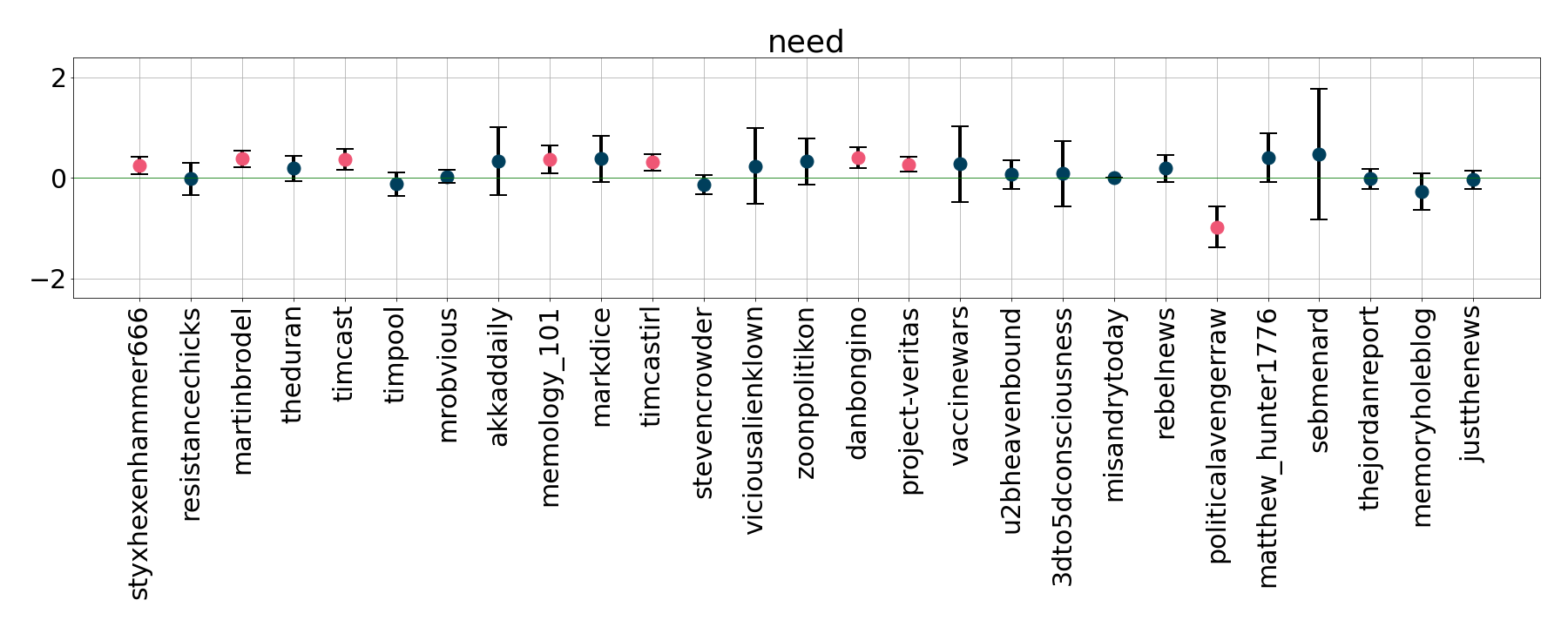}&
\includegraphics[width=.5\textwidth,valign=m]{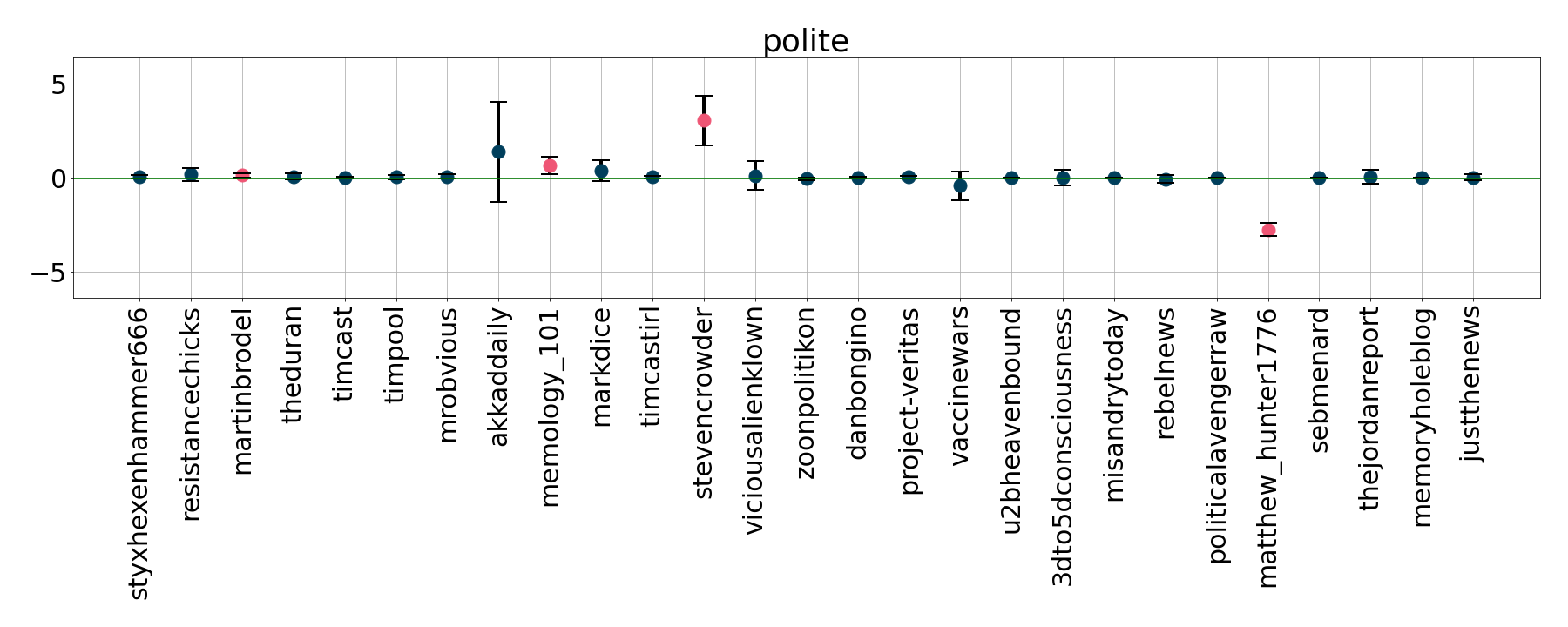}\\
\includegraphics[width=.5\textwidth,valign=m]{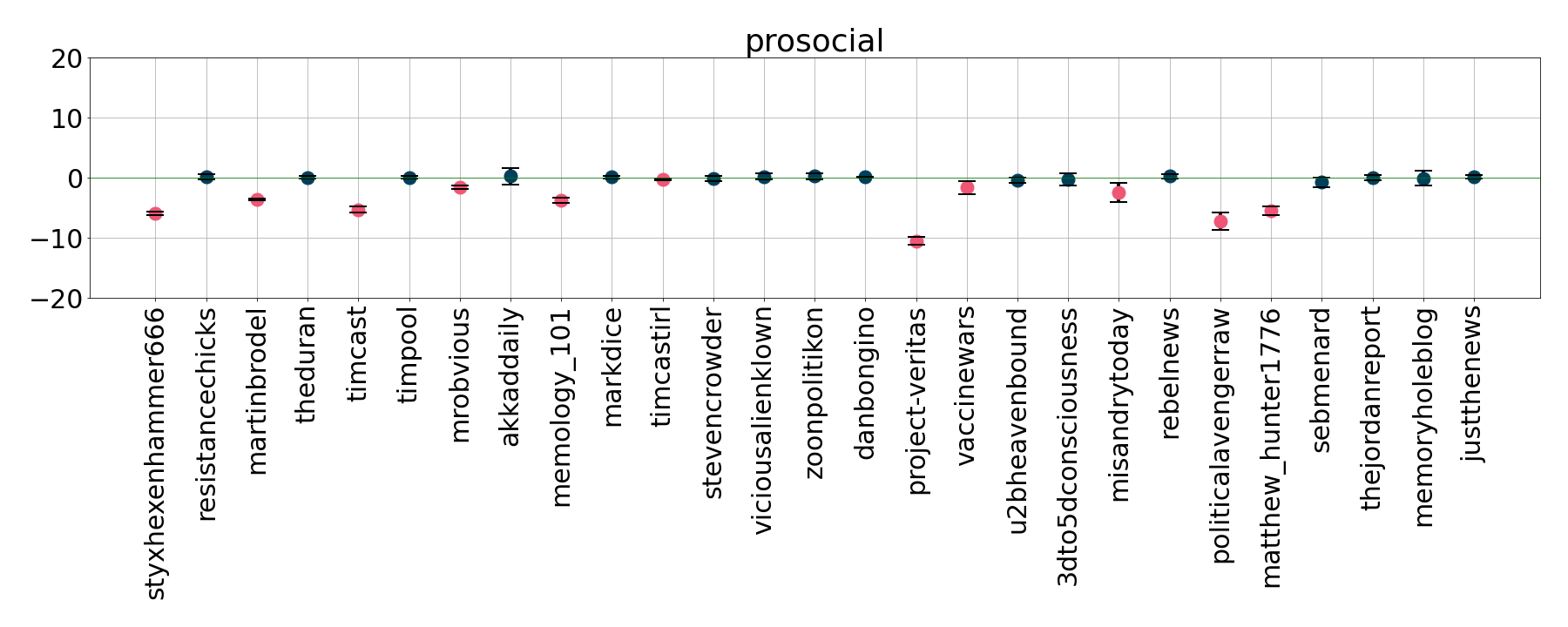}&
\includegraphics[width=.5\textwidth,valign=m]{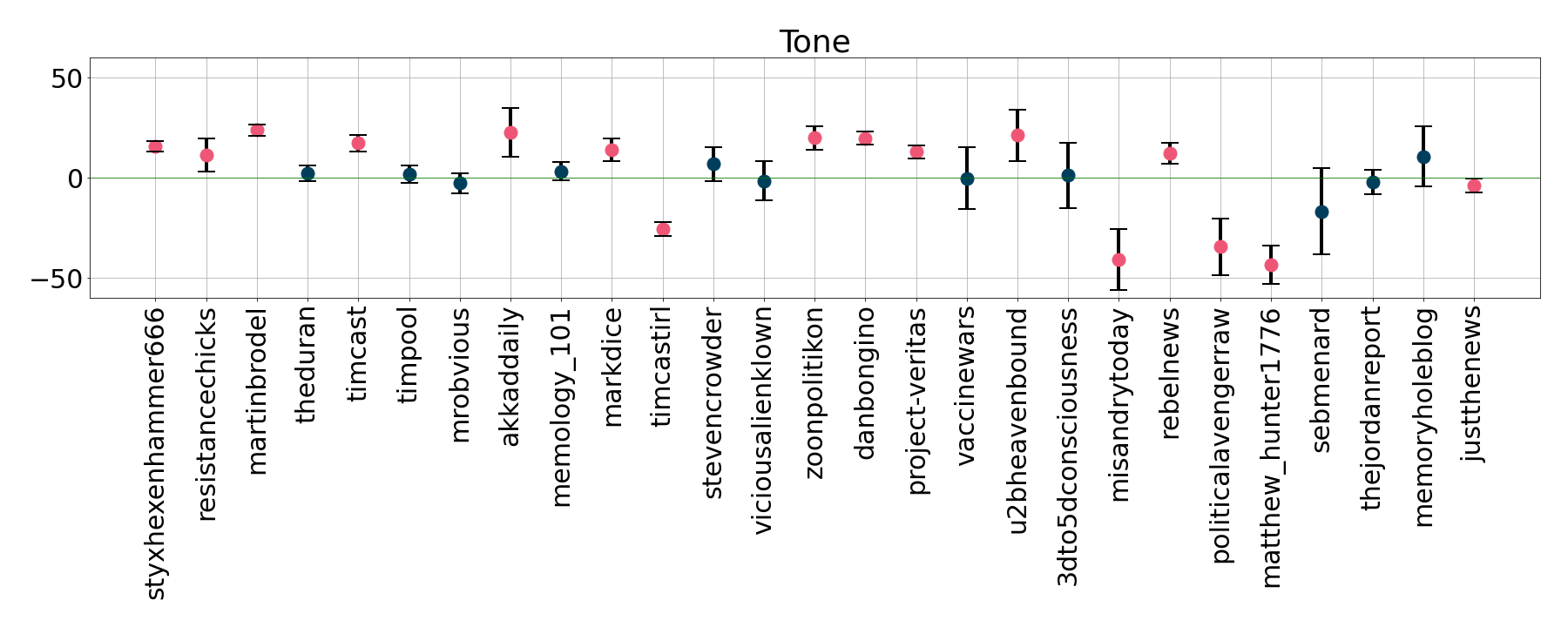}\\
\multicolumn{2}{c}{\includegraphics[width=.11\linewidth, valign=m]{figures/legends/feature_legend.png}}\\
\end{tabular}
\captionof{figure}{Channel-by-channel differences across platforms of selected LIWC features, where negative values indicates the feature is greater on YouTube, positive values indicate the feature is greater on BitChute, and zero indicates no difference across platforms. If those differences are significant according to Tukey’s HSD, they are highlighted in red.}
\label{featfigs2}
\end{table*}

\begin{small}
\bibliography{scibib}
\end{small}

\end{document}